\documentclass[useAMS,usenatbibi,usegraphicx]{mn2e}
\usepackage{psfig}
\usepackage{latexsym}
\usepackage{amssymb}
\usepackage{amsmath}

\def\kms{\,{\rm km\,s^{-1}}}

\def\msun{\,{\rm M_\odot}}

\def\etal{{et al.\ }}

\newcommand\beq{\begin{equation}}
\newcommand\eeq{\end{equation}}
\newcommand{\ba}{\begin{eqnarray}}
\newcommand{\ea}{\end{eqnarray}}
\def\spose#1{\hbox to 0pt{#1\hss}}
\def\lta{\mathrel{\spose{\lower 3pt\hbox{$\mathchar"218$}}
     \raise 2.0pt\hbox{$\mathchar"13C$}}}
\def\gta{\mathrel{\spose{\lower 3pt\hbox{$\mathchar"218$}}
     \raise 2.0pt\hbox{$\mathchar"13E$}}}

\title[Hypervelocity stars]
{Hypervelocity stars and the environment of Sgr A$^*$}

\author[Sesana, Haardt, \& Madau]
{Alberto Sesana$^{1,2}$, Francesco Haardt$^{1}$,  \& Piero Madau$^{3,4}$ \\
$^1$Dipartimento di Fisica e Matematica, Universit\'a dell'Insubria, Via Valleggio 11, 
22100 Como, Italy\\
$^{2}$School of Physics and Astornomy, University of Birmingham, 
Edgbaston, Birmingham, B15 2TT, UK\\
$^3$Department of Astronomy \& Astrophysics, University of California, Santa Cruz, CA 
95064, USA\\
$^4$Max-Planck-Institut fuer Astrophysik, Karl-Schwarzschild-Strasse 1, 85740 Garching 
bei Muenchen, Germany
}

\begin{document}

\date{Received ---}

\maketitle

\begin{abstract}
Hypervelocity stars (HVSs) are a natural consequence 
of the presence of a massive nuclear black hole (Sgr A$^*$) in 
the Galactic Center. Here we use the Brown et al. sample of unbound and 
bound HVSs together with numerical simulations of the propagation of HVSs in the
Milky Way halo to constrain three plausible ejection mechanisms: 1) the scattering of stars bound 
to Sgr A$^*$ by an inspiraling intermediate-mass black hole (IMBH); 2) the disruption of stellar 
binaries in the tidal field of Sgr A$^*$; and 3) the two-body scattering of stars off 
a cluster of stellar-mass black holes orbiting Sgr A$^*$. We compare the predicted 
radial and velocity distributions of HVSs with the limited-statistics dataset currently 
available, and show that the IMBH model appears to produce a spectrum of ejection velocities that is
too flat. Future astrometric and deep wide-field surveys of HVSs should shed 
unambiguous light on the stellar ejection mechanism and probe the Milky Way potential on 
scales as large as 200 kpc. 
\end{abstract}

\section{Introduction}
Hypervelocity stars (HVSs), i.e. stars moving with speeds sufficient to escape the 
gravitational field of the Milky Way (MW), were first recognized by Hills (1988) as an 
unavoidable byproduct of the presence of a massive black hole in the 
Galactic Center (GC). We now know of seven HVSs in the MW halo traveling with 
Galactic rest-frame velocities $v_{\rm RF}$ in the range between $+400$ and $+750\,\kms$ 
(Brown et al. 2005, 2006a,b; Hirsch et al. 2005; Edelmann et al. 2005). 
Most are probably B-type main sequence halo stars with galactocentric distances of 
50-100 kpc, and have travel times from the GC consistent with their 
lifetimes. Only a close encounter with a relativistic potential well can 
accelerate a 3-4 $\msun$ star to such extreme velocities, and at least three 
different ejection mechanisms from the dense stellar cusp around Sgr A$^*$, 
the massive black hole in the GC, have been proposed: 
\begin{enumerate} 

\item[(1)] the scattering of stars bound to Sgr A$^*$ by an inspiraling intermediate-mass black 
hole (``IMBH model'', Yu \& Tremaine 2003; Levin 2006; Baumgardt et al. 2006; Sesana et al. 2006, 2007b). 

\item[(2)] the tidal breakup of a tight stellar binary by Sgr A$^*$ 
(hereinafter the ``TB model''). This interaction leads to the capture of one 
star and the high-speed ejection of its companion (Hills 1988; Yu \& Tremaine
2003; Gualandris etal. 2005; Ginsburg \& Loeb 2006; Bromley et al. 2006). 

\item[(3)] the scattering of ambient stars by a cluster of stellar-mass black holes 
that have segregated around Sgr A$^*$ (O'Leary \& Loeb 2006; Miralda-Escud\'e \& Gould 2000;
hereinafter the ``BHC'' model).

\end{enumerate} 
In theory, the observed frequency, spectral properties, and spatial and velocity distributions
of HVSs should all shed light on the ejection mechanism and the stellar environment around Sgr A$^*$.
In practice, however, different scenarios can reproduce the inferred rate of removal from the 
GC simply by changing, within the observational constraints, the stellar mass 
function and/or the fraction of stellar binaries. 
Travel times estimates for the known HVSs are spread uniformly between 
30 and 160 Myr, and there is as yet no evidence for a burst of HVSs from the GC (Brown et al. 2006a). 
Both models TB and BHC predict HVSs to be expelled isotropically at an approximately 
constant rate, while in model IMBH HVSs are ejected preferentially within the orbital plane of 
the black hole pair in a short burst lasting a few Myr (Levin 2006; Sesana et al. 2006, 2007a, 2007b). 
Even in the latter case, however, the observed HVS population would plausibly be produced 
by a series of IMBH inspiral events (at a rate that could be as high as 10$^{-7}$ yr$^{-1}$, 
see Portegies Zwart et al. 2006) with randomly oriented orbital planes. HVSs would then 
be distributed isotropically in the halo of the MW, and the imprint of a single burst on their 
spatial distribution would be hardly recognizable. 

In this {\it Letter} we use numerical simulations of the propagation of HVSs in the Milky Way
halo to compare the radial and velocity distributions predicted by the three models to the 
Brown \etal (2006a,b) sample of unbound and bound HVSs.

\begin{figure}
\centerline{\psfig{figure=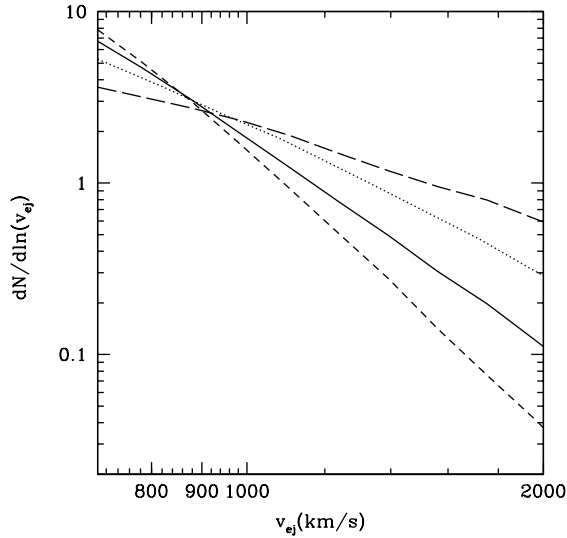,width=3in}}
\caption {Distribution of ejection velocities of HVSs predicted by 
the three different mechanisms discussed in the text.  
{\it Short-dashed line}: model TBln. 
{\it Solid line}: model TBf. 
{\it Dotted line}: model BHC.  
{\it Long-dashed line}: 
model IMBH with $q=1/729$ and $e=0.9$. 
Models TBf and TBln assume a Salpeter 
initial mass function from 1 to 15 $\msun$.  Only stars with ejection 
velocities $v_{\rm ej}> 700\,\kms$ will reach galactocentric distances $r>10$ kpc. 
The average ejection velocities in the plotted range are 870, 930, 990, and 1160 $\kms$
for models TBln, TBf, BHC, and IMBH, respecively. 
Curves are normalized so that 
the integral in $d\ln(v_{\rm ej})$ from $\ln(700/\kms)$ to infinity is unitary.
}
\label{fig1}
\end{figure}

\section{Models}

HVSs are assumed to be ejected from the GC at a steady, arbitrary rate, and their
velocities are obtained as follows:

\begin{enumerate}

\item[(1)] In model IMBH, HVSs are produced according to the velocity distribution 
of scattered stars found in scattering experiments (Sesana et al. 2006, 2007b).
In Sesana et al. 2007b, we study the inspiral of an IMBH onto a MBH surrounded
by a cusp of bound stars. 
Motivated by recent N-body simulations (Matsubayashi et al. 2007), 
we assume that the IMBH starts to eject
stars when the total stellar mass inside the binary semimajor axis $a$ is $\sim M_2$, the 
mass of the secondary.
For Sgr A$^*$, this translates into $a \simeq 0.03$ pc.
We further assume a black hole binary mass ratio $q=1/729$, an initial orbital 
eccentricity $e=0.9$, and a stellar cusp (bound to Sgr A$^*$) with density 
profile $\propto r^{-1.5}$. 
The IMBH is found to stall after loss--cone depletion at $a \sim 0.004$ pc.
While the ejection rate depends on the values of $q$ and $e$, neither the binary 
mass ratio nor its eccentricity have a large effect on the average ejection velocity
(see also Fig. 6 of Sesana \etal 2006).  We checked that either assuming 
$q=1/243$ or $e=0.1$, does not affect significantly the predicted velocity distribution.
As the ejection velocity in a scattering event is a function of binary separation,
the velocity distribution used here is averaged during the entire shrinking phase
of the binary: the IMBH decays fast at large separations (where stars 
gain relatively little energy after an encounter), and ejects stars at higher and
higher speeds as it approaches the hardening radius.  

\item[(2)] For models TB invoking the tidal break-up of stellar binaries by 
a close encounter with Sgr A$^*$, we use the results of the scattering experiments 
performed by Bromley \etal (2006). There, randomly oriented circular binaries are 
launched towards Sgr A$^*$ from a distance of several thousands AU at an initial 
approach speed of 250 $\kms$. The high-speed ejection of a binary member 
depends on the binary semi-major axis $a_{\rm bin}$, the closest approach distance 
between the binary and the hole, $R_{\rm min}$, and the masses of the three bodies. 
A Gaussian distribution of ejection speeds with 20\% dispersion around the mean 
provides a reasonable characterization of the numerical results. We use this
simple Gaussian model, assuming both a flat distribution in log $a_{\rm bin}$ 
(Heacox 1998, hereinafter TBf model), and a lognormal distribution in $a_{\rm bin}$ 
(Duquennoy \& Major 1991, hereinafter TBln model). We randomly sample the closest approach 
distance $R_{\rm min}$ between 1 and 700 AU, and neglect any preference in the 
ejection of either members of the binary. Both binary member masses are generated
according to a Salpeter initial mass function in the range 1-15 $\msun$. 

\item[(3)] Finally, in model BHC, HVSs are generated from the conservative distribution 
of O'Leary \& Loeb (2006, see their Figure 1), including encounters that results
in physical star-black holes collisions. This is because the typical relative 
speed of the stars and the black holes is much larger than the surface escape
velocity of the star: such encounters do not lead to coalescences and could also
result in HVSs. All black holes have a mass of $m_{\rm BH}=10\,\kms$, are 
distributed isotropically, and follow a cuspy density profile with 
slope $-2$ (O'Leary \& Loeb 2006). Compared to model IMBH in which a single scatterer 
slowly sinks inward, in model BHC many scattering centers are present at any given 
time around Sgr A$^*$.

\end{enumerate}

The distribution of stellar ejection velocities at infinite distance from Sgr A$^*$
(and in the absence of other gravitational sources) is shown in Figure \ref{fig1} 
for the three scenarios discussed above, in the range $700<v_{\rm ej}<2000\,\kms$. 
Mechanism IMBH clearly produces a more numerous population of high-speed events with 
$v_{\rm ej}>900\,\kms$ compared to mechanisms TB and BHC. Note that, while in both 
models IMBH and BHC the ejection velocity is independent of stellar mass 
(for model BHC this is actually true only as long as the star is lighter than the
scattering hole), in model TB the ejection velocity of the primary (secondary) component
of a $m_1>m_2$ stellar binary scales as 
$\sqrt{m_2}(m_1+m_2)^{-1/6}$ [$\sqrt{m_1}(m_1+m_2)^{-1/6}$] (Hills 1988).

\section{Simulations}

To generate a simulated catalog of HVSs in the MW halo, we sample the distributions in 
Figure \ref{fig1} and integrate the orbits of ejected stars in a spherically symmetric 
potential using the fourth-order Runge-Kutta routine DOPRI5 (Dormand \& Prince 1978). 
The fractional tolerated error, in position and
velocity of the star, is set to $10^{-12}$ per step, allowing a (fractional) total energy conservation
accuracy $\sim 10^{-8}$. 

The Galactic potential, the main-sequence lifetime of the stars $t_{\rm ms}$, and 
its ejection time relative to the present  all determine the distribution of observable  
velocities as a function of galactocentric distance.
For each star we randomly generate a time $T$ since ejection between zero 
and the main sequence lifetime $t_{\rm ms}$, integrate its orbit for a time 
$T$, and then store its final distance and velocity. While stars ejected with lower 
speeds cannot reach large distances within a time $<t_{\rm ms}$ and will only 
populate the inner halo, the high-velocity tail can reach more distant regions of the 
MW within the stellar main-sequence lifetime (Bromley \etal 2006). 

Stars are assumed to be injected in the GC at a constant rate according to a Salpeter IMF,
$dN/dm_*\propto m_*^{-2.35}$. The scattering rate of stars having mass between 
$m_*$ and $m_*+\Delta m_*$ is then given by
$
t_c^{-1}\,(dN/dm_*)\,\Delta m_*,
$
where $t_c$ is the characteristic timescale between encounters. Since
the orbits of stars of mass $m_*$ are only followed for at most a
main-sequence timescale $t_{\rm ms}(m_*)$, we must normalize the total number
of events according to
\begin{equation}
\Delta N(m_*)=\frac{t_{\rm ms}(m_*)}{t_c}\,\frac{dN}{dm_*}\,\Delta m_*.
\label{eq:dndm}
\end{equation}
The number of stars ejected in the same mass interval is then $\Delta N(m_*)F(m_*)$, 
where the dimensionless function $F(m_*)$ takes into account the mass dependence of 
the ejection mechanism (as in model TB the ejection probability is larger for low-mass 
binaries, Hills 1988).
Note that, in all models, the stellar lifetime introduces an explicit dependence on 
stellar mass in the spatial distribution of observable HVSs, since more massive stars
need higher ejection speeds to reach large galactocentric distances.

To bracket the uncertainties in the MW potential, we have used 5 different models
for the mass distribution in the Galaxy. 
One is the single-component default model used by Bromley et al. 2006 (hereinafter 
potential Bdef), a cored power-law
\begin{equation}
\rho(r)=\rho_0/\left[1+(r/r_c)^2\right], 
\end{equation}
where $\rho_0=1.27\times 10^4\,\msun/{\rm pc}^3$ is the central density, and the core 
radius is $r_c=8$ pc. The other four are multi-component models, formed by a power-law 
stellar bulge, an exponential disk, and an NFW (Navarro, Frank \& White 1997) dark matter 
halo. In model WDa, the disk and halo are chosen according to model MWa of Widrow 
\& Dubinsky (2005), and the bulge mass is set to $1.4\times 10^{10}\msun$. 
In models DB2d and DB4d, disk and halo are chosen according to models 2d and 4d of 
Dehnen \& Binney (1998), and the bulge mass is $0.8\times 10^{10}\msun$. Finally, a
variant of model WDa is constructed where the disk has the same mass but a smaller
scale length $R_d=2$ kpc, and the halo has a larger scale length  
(19 kpc) and is more massive ($M_{h,100}=9.4\times 10^{11}\,\msun$). Such model, 
termed ``Deep", is characterized by a large local escape speed. For simplicity, in all 
models the bulge is spherically symmetric and the disk mass is added as a spherical 
component to the bulge and halo. The disk contribution to the potential is significant 
only in models DB2d and Deep, and only in the inner 3-8 kpc.
Table \ref{tab1} list several observable quantities for the five assumed MW mass 
distributions. The potential gets shallower from top to bottom, with extreme models 
Deep and DB4d bracketing the range allowed by the observations.

\begin{figure}
\centerline{\psfig{figure=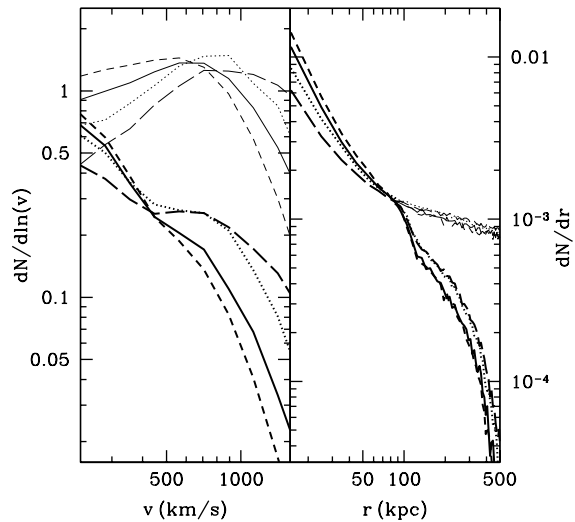,width=3in}}
\caption{Normalized distribution of observable velocities $v>200\,\kms$ and 
galactocentric distances $r>10$ kpc of stars ejected from the GC. Linestyles as 
in Fig. \ref{fig1}. All models assume a Salpeter initial 
mass function in the range $1-15 \msun$, and the MW potential WDa. {\it Thin lines:}
all stars. {\it Thick lines:} stars brighter than $m_V=24.5$ (see \S~4 for details).
{\it Left panel}: velocity distribution. {\it Right panel}: radial distribution.}
\label{fig2amod}
\end{figure}
\begin{figure}
\centerline{\psfig{figure=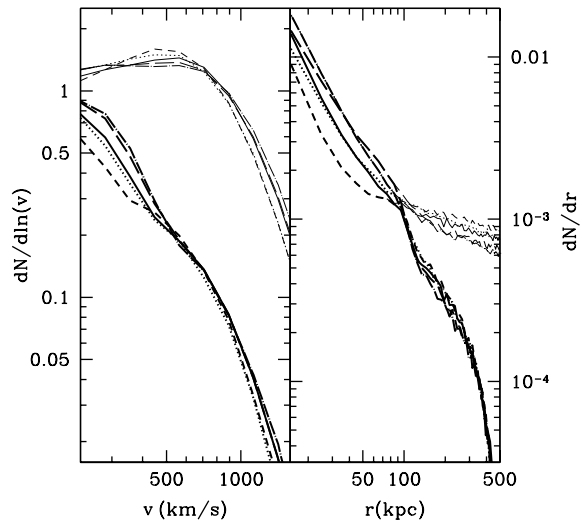,width=3in}}
\caption{
Same as Fig. \ref{fig2amod}, but for the TBln ejection mechanism and different 
MW potentials: 
Deep ({\it dot-dashed lines}), 
Bdef ({\it long-dashed lines}), 
WDa ({\it solid lines}), 
DB2d ({\it dotted lines}) 
and DB4d ({\it short-dashed lines}).} 
\label{fig2apot}
\end{figure}

Combining the different ejection mechanisms with the MW gravitational potentials 
yields a total of twenty different models for the distributions observable speeds and
distances of HVSs. These are shown in Figures \ref{fig2amod} and \ref{fig2apot}
for stars with $v>200\,\kms$ and $r>10$ kpc. The velocity distribution shows a 
broad peak for $v\sim 500-800\,\kms$, and different ejection scenarios are clearly recognizable.
For the same ejection mechanism, the effect of the different MW potentials is relatively
weak (Fig. \ref{fig2apot}, left panel). The distribution of galactocentric distances 
is instead quite sensitive to both the ejection and the potential model, particularly
for stars within 100 kpc of the GC. Note that a steady rate of ejection of HVSs from the 
GC would result in a flat $dN/dr$ curve. The excess at $r \lta 50$ kpc is due to the many 
low-mass stars that are scattered into the halo on bound orbits.

\begin{table}
\begin{center}
\begin{tabular}{lccccccc}
\hline
$$  & $M_{0.01}$ & $M_{100}$ & $V_{c,8}$ & $A$ & $B$ &  $V_{\rm esc}$ & $V_{\rm esc,8}$\\
\hline
 Deep&  3.46&   10.0& 207&    15.0&     -11.0&     978&   630\\
 Bdef&  2.93&     10.2& 210&   13.2&      -13.2&     929&   585\\
 WDa&   3.45&     6.7& 229&   14.8&       -13.9&      899&   550\\
 DB2d&  2.98&     6.3& 206&   14.0&      -11.9&       824&   515\\ 
 DB4d&  2.97&     4.0& 217&   14.4&      -13.0&       791&   455\\
\hline
\end{tabular}
\end{center}
\caption{The five different models of MW mass distribution discussed in the text.
The quantities $M_{0.01}, M_{100}, V_{c,8}, A, B, 
V_{\rm esc},$ and $V_{\rm esc,8}$ are, respectively, the mass enclosed in 10 pc 
in units of $10^7\msun$, the mass enclosed in 100 kpc in units of $10^{11}\msun$,
the circular velocity at 8 kpc in $\kms$, the Oort constants, the escape velocity 
from Sgr A$^*$ in $\kms$; the escape velocity at 8 kpc in $\kms$.
}
\label{tab1}
\end{table}

\section{Comparison with the observations}

\begin{figure}
\centerline{\psfig{figure=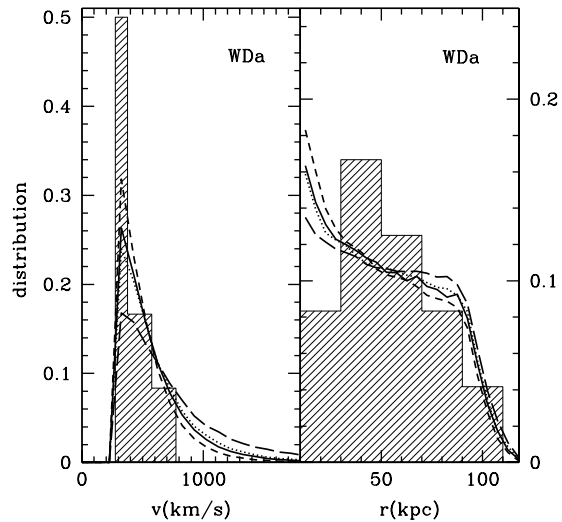,width=3in}}
\caption{Theoretical distributions of observable speeds ({\it left panel}) and 
galactocentric distances ({\it right panel}) of ejected stars  
in the WDa potential. Model predictions are shown for stars in the mass range 3-4 $\msun$,
distances $r>10$ kpc, and velocities $v>+275\,\kms$. Linestyles as in Fig.  
\ref{fig1}. The shaded histogram represents one possible realization from the Brown \etal 
(2007) sample of bound and unbound HVSs.}
\label{fig2}
\end{figure}

We compare our predicted observable distributions to the Brown et al. sample 
of HVSs. The sample comprises 5 stars with $v_{\rm RF}>+400\,\kms$ and $r>50$ kpc 
(the unbound genuine HVSs of Brown \etal 2006a), as well as 7 candidate HVSs with 
$+275<v_{\rm RF}<+450\,\kms$ and distances 
greater than 10 kpc (the new class of possible ``bound HVSs" recently advocated by 
Brown \etal 2007). 
The combined survey selects B stars with $3\msun<m_*<4\msun$ down to a faint
magnitude limit of $m_V=19.5$, and is 100 \% complete across the high declination region of the Sloan Digital Sky Survey, Data Release 4.
Completeness is to a depth of 10-120 kpc (10-90 kpc) for 4 (3) $\msun$ stars.
As the actual ejection rates of the different mechanisms 
depend upon the poorly constrained properties of the stellar population in the GC
(and, for model IMBH, on the eccentricity and mass of the secondary black hole),
we only consider here normalized distribution of velocities and distances. 
Model predictions are shown only for stars in the mass range 3 to 4 $\msun$.  
The theoretical distributions of observable velocities at all distances $> 10$ kpc are 
plotted in Figure~\ref{fig2}
against the observations. Model IMBH clearly produces a long tail of HVSs with $v>1500\,\kms$, and 
a mean velocity of ejected stars that is larger compared to other mechanisms. For example,
the mean velocity of HVSs in a WDa potential is $(594, 429, 403, 343)\,\kms$ for models IMBH, 
BHC, TBf, and TBln, respectively. The TBf and BHC scenarios predict similar distributions, 
with BHC slightly shifted towards higher velocities. 
Predictions for the tidal break-up models are somewhat sensitive to the
semimajor axis distribution of stellar binaries. 
In model TBln close binaries are rarer compared to model TBf, and the high-velocity 
tail of the velocity distribution is less pronounced. 
Table \ref{tab2} shows the results of the two-dimensional, two-sample Kolmogorov-Smirnov 
(KS) test (Press et al. 1992) applied to the observed bivariate distance/velocity 
distribution of HVSs. This is compared to synthetic data sets of the same size drawn 
from the theoretical distributions of HVSs with masses of 3-4 $\msun$, distances $r>10$ 
kpc, and velocities $v>+275\,\kms$. 
\footnote{Note that the analysis of Brown \etal (2007) yields a statistically significant
excess of 7 out of 11 observed ``bound" stars, allowing for a total of 330 random 
realizations of the real data.} Statistically, there is no difference between data and 
simulations only in the case of model TBln in shallow potentials (DB2d and DB4d). 
Within the limited-statistics dataset currently available, model IMBH appears to be disfavoured.

\begin{table}
\begin{center}
\begin{tabular}{lcccc}
\hline
& TBln & TBf & BHC & IMBH\\
\hline
Deep&    0.151&     0.085&       0.057&        0.007\\
  Bdef&    0.150&          0.105&      0.060&        0.010\\
  WDa&     0.165&          0.110&       0.070&       0.008\\
  DB2d&    0.215&          0.149&       0.127&        0.014\\
  DB4d&    0.209&          0.147&       0.127&        0.015\\
\hline
\end{tabular}
\end{center}
\caption{Two-dimensional, two-sample Kolmogorov-Smirnov test significance for the observed 
distance-velocity distribution of HVSs compared to our twenty different (ejection mechanisms/MW 
potential) models.}
\label{tab2}
\end{table}

\begin{figure}
\centerline{\psfig{figure=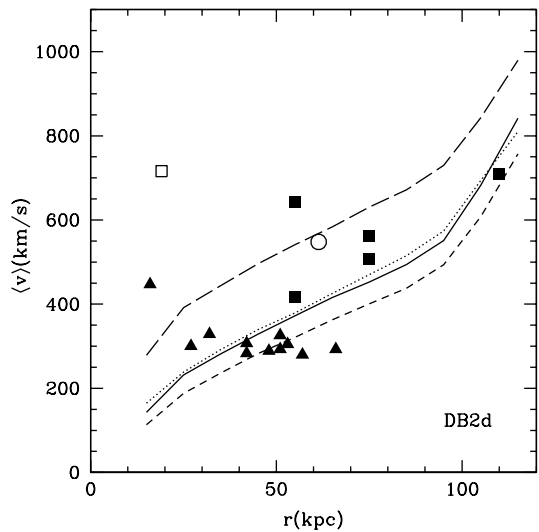,width=3in}}
\caption{Mean observable speed of ejected stars in the mass range 3-4 $\msun$, 
as a function of galactocentric distance, for a DB2d MW potential. 
Linestyles as in Fig. 1.
{\it Filled squares}: the 5 genuine unbound HVSs of the Brown et al. (2006a)
sample.  {\it Filled triangles}: the 11 stars with $+275<v_{\rm RF}<+450\,\kms$ of
the Brown \etal sample. About 7 of them are candidate bound HVSs. 
{\it Open square}: the HVS from Hirsch et al. (2005).  
{\it Open circle}: the HVS from Edelmann et al. (2005).
}
\label{fig4}
\end{figure}

Slowly-moving, short-lived massive stars can only populate the inner halo, and ejection 
mechanisms like TBln and TBf that produce many low-speed events will generate 
a spatial distribution with an ``excess" of stars within $\sim$50 kpc of the GC.  
Figure \ref{fig4} shows the mean observable speed of 3-4 $\msun$ HVSs as a function 
of galactocentric distance. This is an increasing function of $r$ because of stellar lifetime 
effects. Model IMBH appears to produce an excess of extremely fast-moving stars compared to the 
observations. We stress that this result does not depend on the values assumed
for $q$ and $e$. In particular, we tested that increasing $q$ and/or decreasing $e$ 
do not suppress the high velocity tail.

\section{Discussion}

HVSs are in principle a powerful probe of the MW dark matter halo. Table~\ref{tab2} 
shows that, for any given ejection mechanism, the currently available 
statistics is far too low to constrain the MW potential (cf. Bromley \etal 2006).  
The situation will change dramatically with future astrometric (Gnedin \etal 2005) and deep 
wide-field surveys. For illustrative purposes, we focus here on {\it Large Synoptic 
Survey Telescope} ({\it LSST}, Claver et al. 2004). The {\it LSST} is being designed 
to survey an area of 20,000 deg$^2$ with 15 secs pointings, down to a  limiting magnitude 
of $m_V=24.5$. To assess the impact of the {\it LSST} on studies of HVSs,
we first assign a survey volume to the theoretical models discussed in the previous 
sections and use the Brown et al. sample to estimate the expected number of   
HVSs detectable by the {\it LSST}. Assuming that the model distributions do indeed describe 
the parent population of HVSs, we then extract randomly from each of them a mock
catalogue of HVSs. We then apply K-S statistics to quantify the ability of the {\it LSST} 
to distinguish between different scenarios.  

As an illustrative example, we show in Figure \ref{fig2amod} the distribution of 
observable speeds and distances predicted by the different ejection scenarios,
for all $v>+200\,\kms$ and $r>10$ kpc HVSs in the {\it LSST} survey volume. 
At a magnitude limit of $m_V=24.5$, a $1\,\msun$ star can be detected at a
distance of $\lta 100$ kpc, and the distance distributions drop rapidly at larger distances. 
Assuming a Salpeter initial mass function 
and ejection model TBln in a DB2d potential, we estimate that the {\it LSST} 
should detect $\sim2500\pm 800$ HVSs with 
$v>+275\,\kms$, in the mass range $1-15\,\msun$. 
We find that the detection of $\gta 100$ HVSs may be 
enough to identify unambiguously the ejection model {\it and} the Galactic potential. 
 

\section*{Acknowledgments}
Support to this work was provided by NASA grants NAG5-11513, NNG04GK85G, 
and by the Alexander von Humboldt Foundation (P.M.).

{}

\end{document}